\documentclass[fleqn,usenatbib]{mnras}

\usepackage{newtxtext,newtxmath}

\usepackage[T1]{fontenc}
\usepackage{ae,aecompl}

\usepackage{graphicx}	
\usepackage{amsmath}	
\usepackage{amssymb}	
\usepackage{subfig}
\usepackage{xcolor}
\hypersetup{draft}

\title[HD163296 submillimetre dust polarisation and opacity]{Submillimetre dust polarisation and opacity in the HD163296 protoplanetary ring system}

\author[W.R.F.Dent et al.]{
W. R. F. Dent$^{1}$\thanks{E-mail: wdent@alma.cl},
C. Pinte$^{2,3}$,
P. C. Cortes$^{1,4}$,
F. M\'enard$^{3}$,
A. Hales$^{1,4}$,
\newauthor
E. Fomalont$^{1,4}$,
I. de Gregorio-Monsalvo$^{1}$
\\
$^{1}$ALMA JAO, Alonso de Cordova 3107, Santiago, Chile\\
$^{2}$Monash Centre for Astrophysics (MoCA) and School of Physics and Astronomy, Monash University, Clayton Vic 3800, Australia\\
$^{3}$Univ. Grenoble Alpes, CNRS, IPAG, F-38000 Grenoble, France\\
$^{4}$National Radio Astronomy Observatory, 520 Edgemont Road, Charlottesville, VA 22903-2475, United States of America\\
}

\date{Accepted XXX. Received YYY; in original form ZZZ}

\pubyear{2018}

\begin{document}
\label{firstpage}
\pagerange{\pageref{firstpage}--\pageref{lastpage}}
\maketitle

\begin{abstract}
We present ALMA images of the sub-mm continuum polarisation and spectral index
of the protoplanetary ringed disk HD163296. The polarisation fraction at 870$\mu m$ is
measured to be $\sim\,$0.9$\%$ in the central core and generally increases with
radius along the disk major axis. It peaks in the gaps between the dust rings, and
the largest value ($\sim\,$4$\%$) is found between rings 1 and 2. The polarisation
vectors are aligned with the disk minor axis in the central core, but become more azimuthal in the gaps, twisting by up to $\pm
9\degr$ in the gap between rings 1 and 2. These general characteristics are consistent with a model
of self-scattered radiation in the ringed structure, without requiring an additional dust
alignment mechanism. The 870/1300$\mu m$ dust
spectral index exhibits minima in the centre and the inner rings, suggesting these regions
have high optical depths. However, further refinement of the dust or the disk model at higher resolution is needed to
reproduce simultaneously the observed degree of polarisation and the low spectral index.

\end{abstract}




\section{Introduction}

Images of protoplanetary disks, now being routinely obtained on spatial scales of tens of au, are showing that dust frequently lies in concentric rings \citep[e.g.,][]{ALMA2015}. Polarised emission at mm wavelengths is a way to further constrain both the disk physics and dust properties. Originally regarded as a tracer of non-spherical magnetically-aligned grains, it was later recognized that self-scattering of mm radiation can be important; the polarisation could then inform us about  disk geometry, optical depths, and dust geometry \citep{Kataoka2016a, Yang2017}. In the disk mid-plane, collisions with the gas can suppress magnetic field alignment, although non-spherical grains may still be aligned with the dominant local radiation field through radiative torque \citep{Tazaki2017}. The spatial structure of polarised emission is critical to distinguish these mechanisms, but this requires well-resolved images.

The near-face-on transition disk HD142527 showed predominantly radial polarisation vectors consistent with self-scattering, with some outer regions of azimuthal orientation,  possibly from radiatively-aligned grains \citep{Kataoka2016b}.
At 3~mm, the more inclined multi-ringed disk of HL~Tau showed azimuthal polarisation vectors consistent with radiative alignment \citep{Kataoka2017}. However, at 870$\mu m$ the vectors are more parallel and aligned along the minor axis, suggesting that scattering may dominate. Intermediate wavelengths (1.3~mm) show a mixture of the two components \citep{Stephens2017}. The extended full disk IM Lup also shows vectors aligned more with the minor axis, and was interpreted as scattered emission \citep{Hull2018}.
Inclined younger Class~0/I disks also tend to show polarisation vectors aligned with the minor axis \citep{Cox2018,Lee2018,Harris2018}.
\vfill
The target for the present study, HD163296, is a well-studied isolated Herbig Ae star, with a bright, extended disk. Lying at a distance of 105pc\footnote{Sizes and luminosities in this paper have been revised based on the GAIA DR2 parallax}, it is classified as spectral type A2Ve, of age 5Myr \citep{Montesinos2009}. In CO gas, the disk extends to $\sim$450au, but appears only half this size when observed in mm dust \citep{deGregorio2013}. Detailed fits indicated that the dust lies in rings \citep{Zhang2016}, subsequently confirmed by higher resolution mm observations which showed three concentric structures \citep{Isella2016}.  The innermost of these is traced in near-infrared scattered light \citep{Monnier2017,MuroArena2018}. Early sub-mm observations found an upper limit of 1\% to the polarisation fraction in a 1 arcsec beam \citep{Hughes2009}. \cite{Pinte2018} and \cite{Teague2018} have also recently found dynamical evidence of massive planets in the system.

In this paper, we present ALMA observations of polarised dust emission of HD163296 at 870$\mu$m with a resolution of $\sim$0.2$\arcsec$ - sufficient to resolve the rings. We also combine the intensity images with archival band 6 data to obtain spectral index maps. The results are compared with a self-scattering model including the gaps and rings.

\section{Observations}

The observations were conducted in sessions on two nights: 23 Jul and 13 Aug 2016, with two consecutive executions of the scheduling block on each night and 35-37 antennas in the array. The polarisation calibrator (J1751+0939) was observed every 30 minutes; J1742-2527 was used as a complex gain calibrator and observed every 7 minutes. J1733-1304 was used as a flux calibrator with assumed fluxes of 1.52 and 1.67~Jy at 343.5~GHz on the two observing nights; J1924-2914 was the bandpass calibrator. Checks showed that the independently calibrated fluxes of both the bandpass and polarisation calibrators agreed with the measured values at that time from the ALMA online calibrator database to better than 5\%.  The total time on-source was 2 hours, although some 15 minutes of data were flagged in the second session while the target transited close to the zenith.

The system was set up to observe continuum in band 7 at a mean frequency of 343.5GHz, with spectral windows at 336.494, 338.432, 348.494 and 350.494GHz, each having 31.25MHz channel spacing and 2GHz nominal bandwidth, with a total usable bandwidth of 7.5GHz.

The calibrated data from the ALMA pipeline was examined, further flagging applied, and the individual executions were used to create self-calibrated solutions for the antenna phase and amplitude. The polarisation was then calibrated for each day separately, and the self-cal solutions applied. The two sessions were combined and imaged using \textsc{clean} in CASA \citep{2007ASPC..376..127M}, producing separate images of Stokes I, Q, U and V. These were combined and debiased to give the polarised percentage $P\% = 100\sqrt{{Q}^2 +{U}^2- {rms}^2} / I$, where $rms$ is the noise in the Stokes Q and U images, and $I$ the total intensity. To maximise the signal/noise, natural uv weighting was applied, giving a beam of 0.21x0.19$\arcsec$ at 81$\degr$, and resulting in a final rms of $\sim$50$\mu$Jy in the Stokes images.

To obtain the spectral index, we used the total intensity band 7 visibilities together with ALMA archive data taken in 2015 Aug in band 6 \citep[originally published by ][]{Isella2016}. Data were combined using the multi-frequency synthesis method \citep{Rau2011}, which takes into account differences in the detailed uv coverage (although both datasets have a similar resolution, which minimises such effects). A differential pointing offset of -11, -73 {\it mas} was measured (and corrected for). This may be partly due to the published stellar proper motion of -8, -39 {\it mas}, and partly from ALMA astrometric errors (typically 1/10 of the beam, or 20{\it mas}). The resultant combined total intensity and spectral index images were obtained using \textsc{clean} with Briggs robustness=0.5 uv weighting, to give a slightly higher resolution (0.18x0.17$\arcsec$), and an rms of 55$\mu$Jy at an effective central frequency of 291 GHz.

\section{Results}

\subsection{Polarisation}

Images of the total intensity and polarisation are shown in Fig.\ref{fig:HD_results.png}. These are derived from the individual Stokes datasets (see supplemental online data). The locations of the total intensity rings at radii of 67, 102 and 160~au are illustrated by the three superimposed circles inclined at 45$\degr$ to the line-of-sight. The percentage polarisation (Fig.1b) ranges from 0.9\% in the centre, and increases up to $\sim$4\% at the edges along the major axis; beyond this the signal/noise of the polarised intensity drops below 5$\sigma$ (see Fig.1c), making the fractional polarisation uncertain. In the gaps between rings along the major axis, particularly between ring 1 and 2, the polarised intensity is relatively high compared with the total intensity, resulting in the regions of high fractional polarisation (shown by lighter shading in Fig.1b). This anticorrelation of polarisation fraction and intensity is illustrated further in the major axis cross-cut in the upper panel of Fig.\ref{fig:poln_cut.png}. Note that along the minor axis, the polarised flux is comparatively weak and only marginally detected beyond ring 1 (Fig.1c).

\begin{figure*}
  \includegraphics[width=0.8\columnwidth]{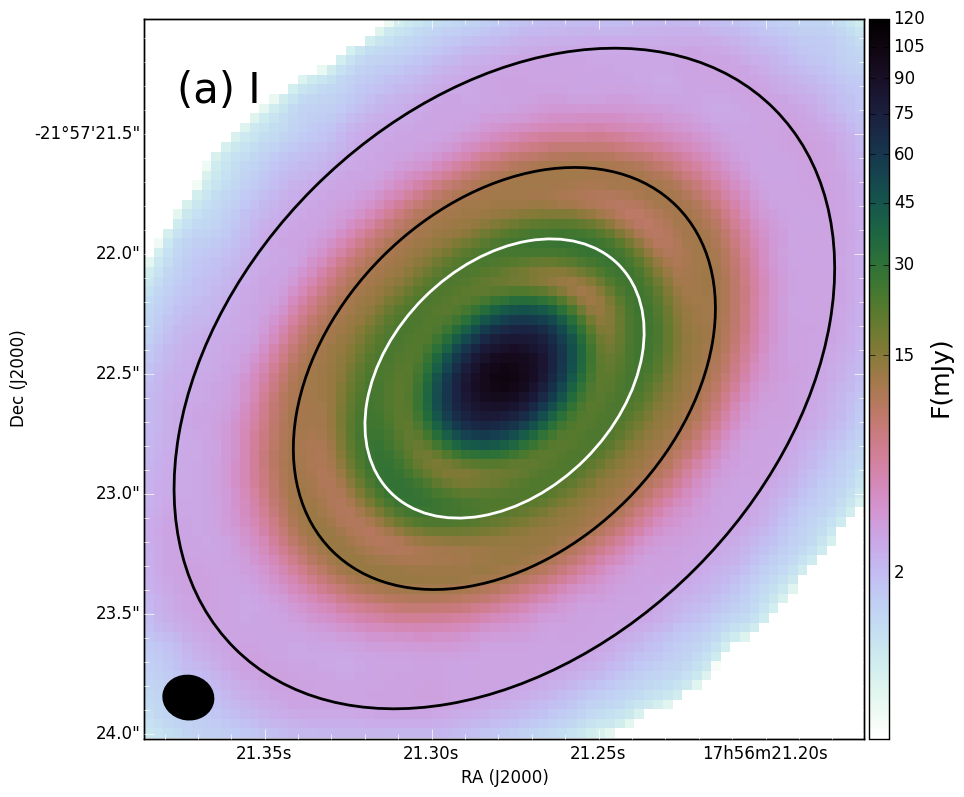}
 \includegraphics[width=0.8\columnwidth]{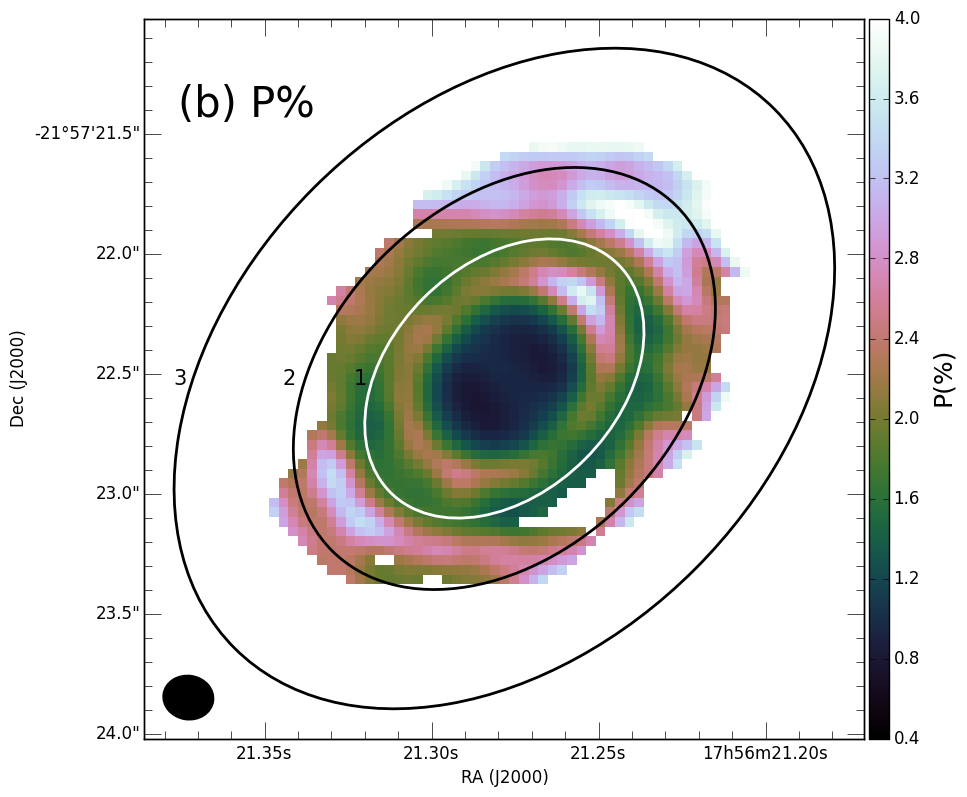}
  \includegraphics[width=0.8\columnwidth]{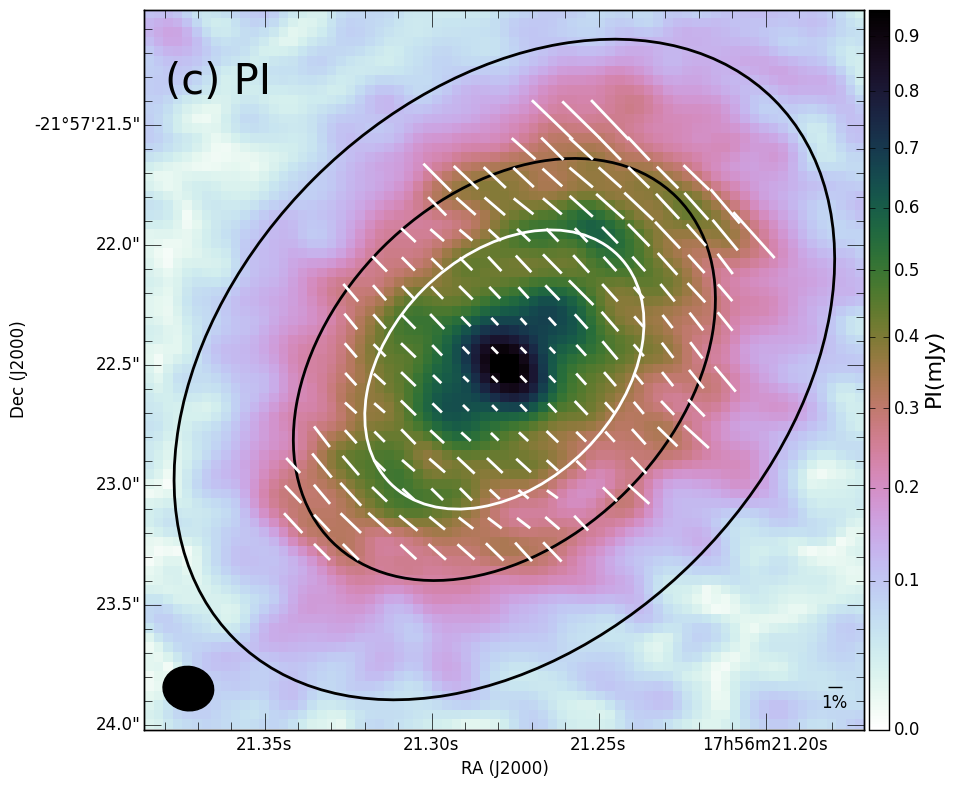}
  \includegraphics[width=0.8\columnwidth]{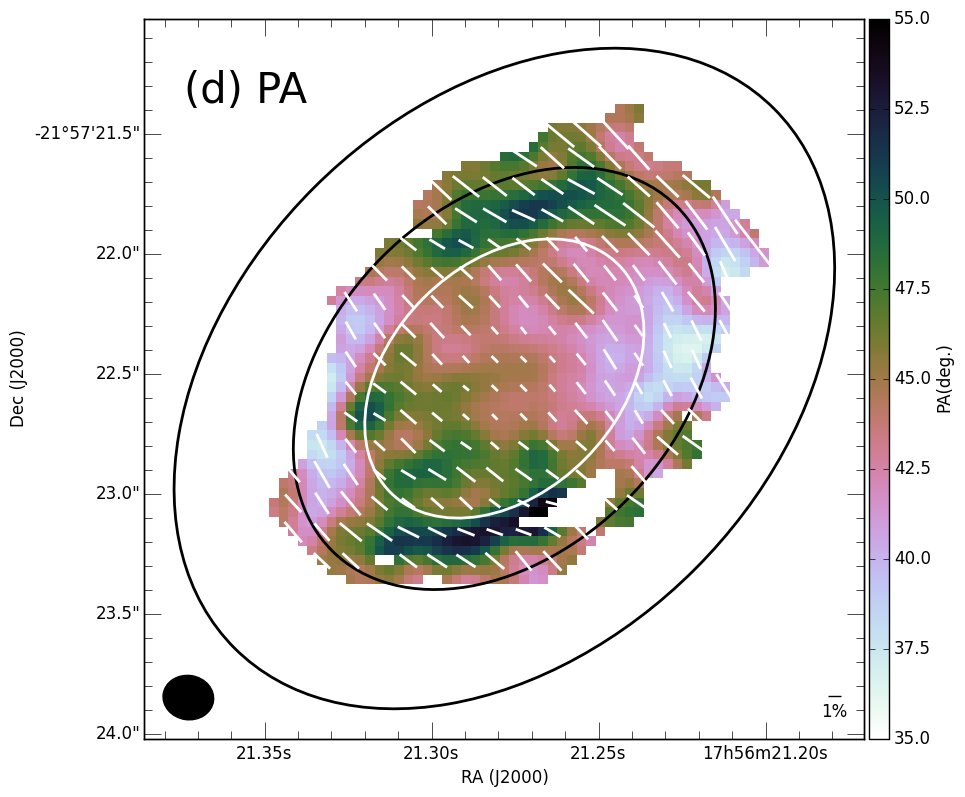}

    \caption{Sub-mm continuum polarisation in HD163296. The top panels show (a) the total intensity and (b) the percentage polarisation at band 7 in the central region. Superimposed as ellipses are representations of the three rings in the total intensity image marked 1 - 3 in (b). Lower panels are (c) the polarised flux intensity and (d) the polarisation angle, both with the polarisation vectors superimposed. In (c), the vectors are shown at their actual angle, whereas in (d) the vector rotation relative to the mean of 44.1$\degr$ is multiplied by a factor of 3, to emphasise the relative twist. Polarisation fraction and angle are truncated at polarised flux levels of 5$\sigma$ per beam (250$\mu Jy$).   Images are 3$\arcsec$ (315~au) across. The images have a resolution of $0.2\arcsec$ (beams are shown lower left). Polarization vectors have 1/2 beam spacing, with lengths proportional to the polarization fraction.}
\label{fig:HD_results.png}
\end{figure*}

Fig.\ref{fig:HD_results.png}c-d show a relatively constant polarisation angle in the central bright region, with a median of 44.1$\pm1.5\degr$ over the inner 30~au - closely aligned with the disk minor axis.
However, there is a twist of up to $\pm~9 \degr$ in the vectors at larger radii, most significant in the intensity gap between ring 1 and 2 (where the fractional polarisation peaks). Here the vectors are more azimuthally aligned; this is more clearly illustrated when the offset in the angle relative to the central polarisation is magnified (see Fig.1d). This azimuthal twist can also be seen at smaller and larger radii in Fig.1d, albeit with a lower magnitude.

\begin{figure}
	\includegraphics[width=1.0\columnwidth]{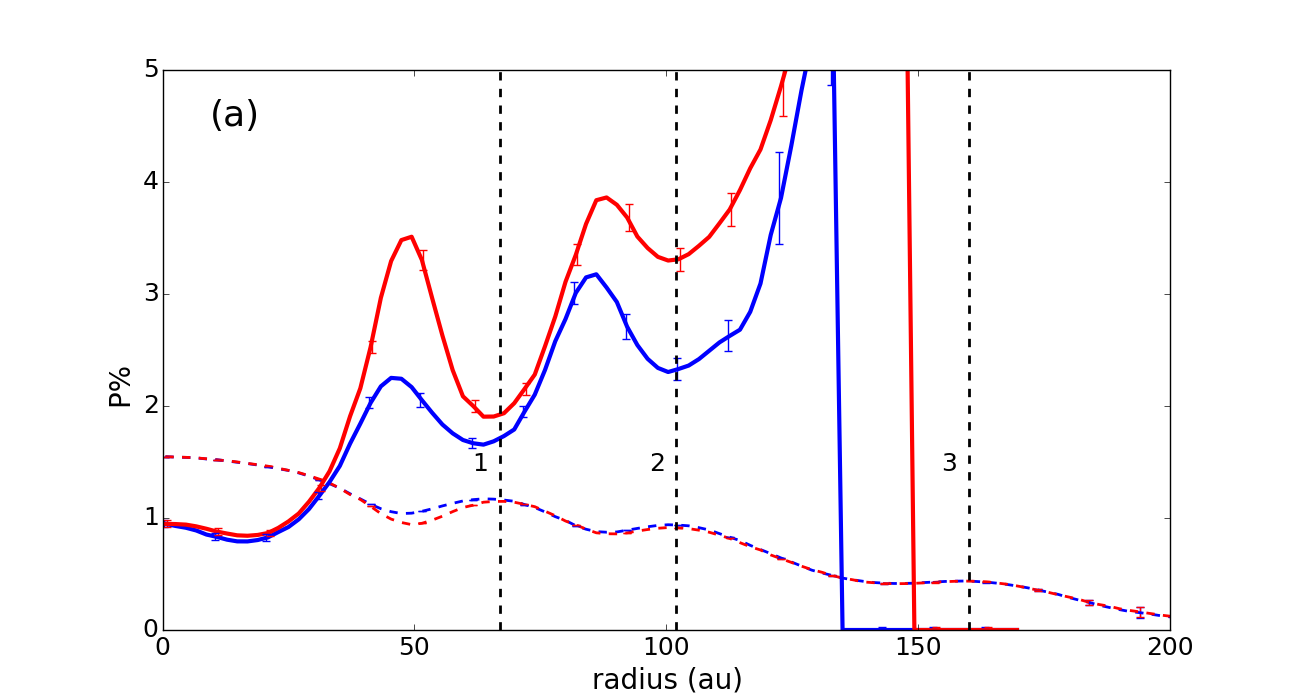}
%
    \hfill
   \includegraphics[width=1.0\columnwidth]{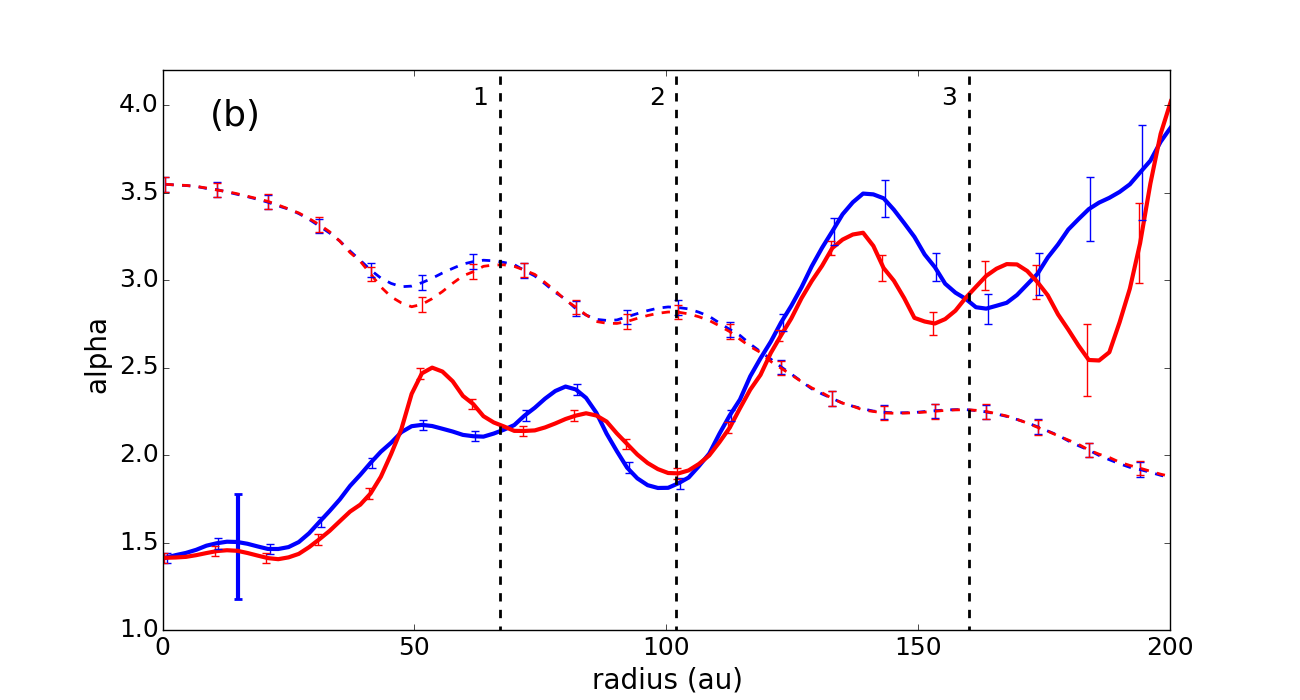}

    \caption{Sector cuts of polarisation percentage (P\%, upper panel) and spectral index $\alpha$ (from band 6+7, lower panel) along the disk major axis. Data are the values along the major axis, averaged over arcs subtending an angle of -45$\pm10\degr$ on the sky (or deprojected $\pm14\degr$ around the disk). Red and blue solid lines represent the NW and SE directions from the central star. For comparison, the dashed lines show the log of total intensity (arbitrary units) from the respective datasets. The radii of the three intensity rings are illustrated by the dotted vertical lines.
Error bars are spaced by approximately one resolution element and depict the errors from random noise and image fidelity. The large error bar at lower left of the lower plot shows the systematic uncertainty in $\alpha$ assuming 7\% calibration errors at the two wavelengths.}
    \label{fig:poln_cut.png}
\end{figure}

\subsection{Spectral index}

\begin{figure}
    \includegraphics[width=1.0\columnwidth]{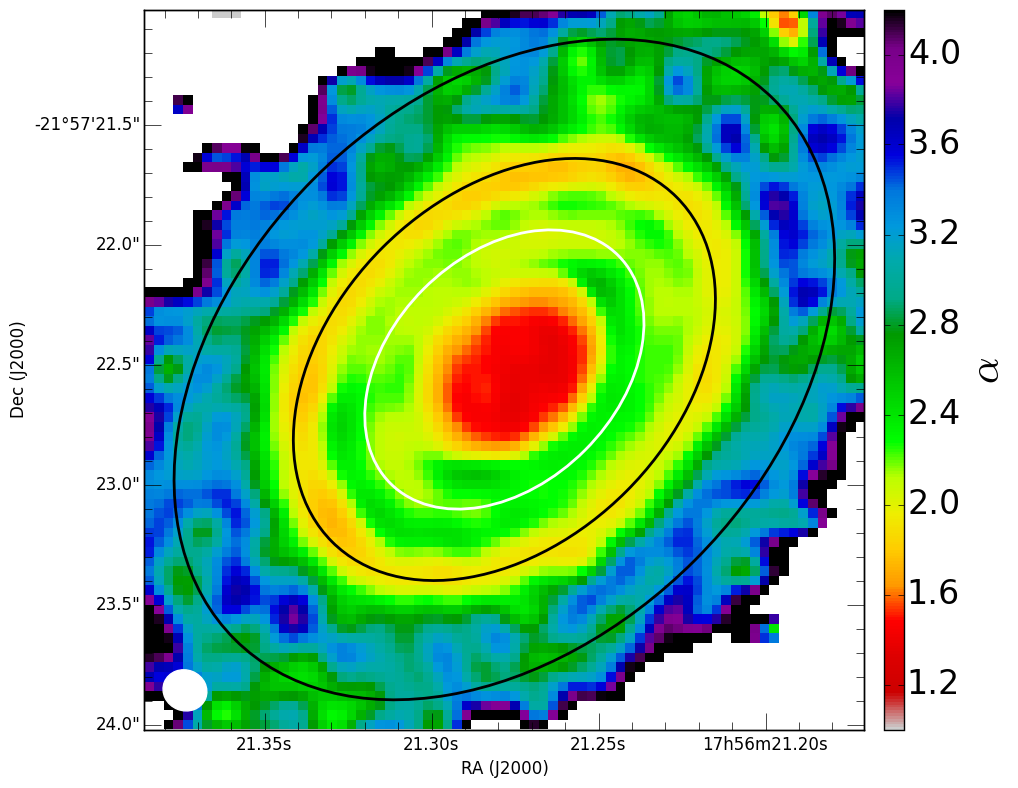}
    \caption{Spectral index map of HD163296. This is derived from the combined band 6 and 7 visibilities, with Briggs 0.5 weighting, resulting in a resolution of $\sim 0.18\arcsec$. The positions of the three bright rings are shown by the ellipses.}
    \label{fig:alpha-image.png}
\end{figure}

The combined band 6+7 image of the spectral index, $\alpha_{mm}$, is shown in Fig.\ref{fig:alpha-image.png},
with a cut along the major axis in Fig.\ref{fig:poln_cut.png}b.  The increase in $\alpha_{mm}$ with radius is consistent with \cite{Guidi2016}, but now we can resolve dips in $\alpha_{mm}$ at the ring peaks. These could be due to a local decrease in the dust emissivity slope $\beta$, or high optical depths in the rings. As their radial width is not resolved, the ring brightness temperature only provides a lower limit to the optical depth, so we cannot discriminate between these possibilities. The rings in HL~Tau \citep{ALMA2015} and TW Hya \citep{Huang2018} also show lower $\alpha_{mm}$, which was associated with high optical depths.

In the central peak of HD163296, $\alpha_{mm}$ is apparently as low as 1.4 - below that of isothermal black body emission in the Rayleigh-Jeans limit (2.0). Contribution from nonthermal emission is considered negligible at these wavelengths \citep{Isella2007}.
The relative flux calibrations across each of the
images are accurate to $\leq1\%$, so the profile of $\alpha_{mm}$ in Fig.2b is well-determined.  However, there may be an
offset that depends on the absolute flux calibration
in the two bands.  A conservative calibration error is
7\%, which would correspond to a spectral index offset of
0.3, illustrated by the large errorbar in
Fig.2b.  So while the relative shape of $\alpha_{mm}$ in Fig.2b
is accurate, the range could change by $\pm0.3$.

The total disk fluxes are 1.68Jy at band 7 and 0.70Jy at band 6, giving $\alpha_{mm}$ of $2.1\pm0.3$. This is in reasonable agreement with that measured by \cite{Pinilla2014} (2.7$\pm0.4$), but suggests an overall offset in $\alpha_{mm}$, marginally consistent with the formal flux uncertainty. Another possibility could be a $\sim20\%$ decrease in luminosity in the interval between the band 6 and 7 observations. Alternatively a hot core surrounded by cool optically thick dust may also reduce $\alpha_{mm}$, as proposed for some Class 0 objects \citep{Li2017}. Interestingly, TW~Hya also showed $\alpha_{mm}<2.0$ in the centre, marginally inconsistent with the calibration accuracy \citep{Huang2018}.

\section{Discussion and modeling}

The mm-wave polarisation vectors in the central region of HD163296 are aligned within 1.5$\degr$ of the disk minor axis. This alignment is similar to that seen in IM Lup \citep{Hull2018}, HL~Tau at 870$\mu$m \citep{Stephens2017} and CW~Tau \citep{Bacciotti2018}, and is consistent with self-scattering in an inclined disk \citep{Kataoka2016a,Yang2016a}.
Emission from non-spherical grains aligned either by magnetic or non-isotropic radiation fields can also give measurable mm polarisation, but for inclined disks like HD163296, they do not predict the same vector orientation along the minor axis in the centre \citep{Yang2016a,Yang2017}.
At larger radii and in the gaps between the rings in HD163296, the polarisation fraction increases and the vectors become more azimuthal. Can this also be explained by self-scattering alone?

\begin{figure*}
  \centering
  \includegraphics[width=0.845
  \linewidth]{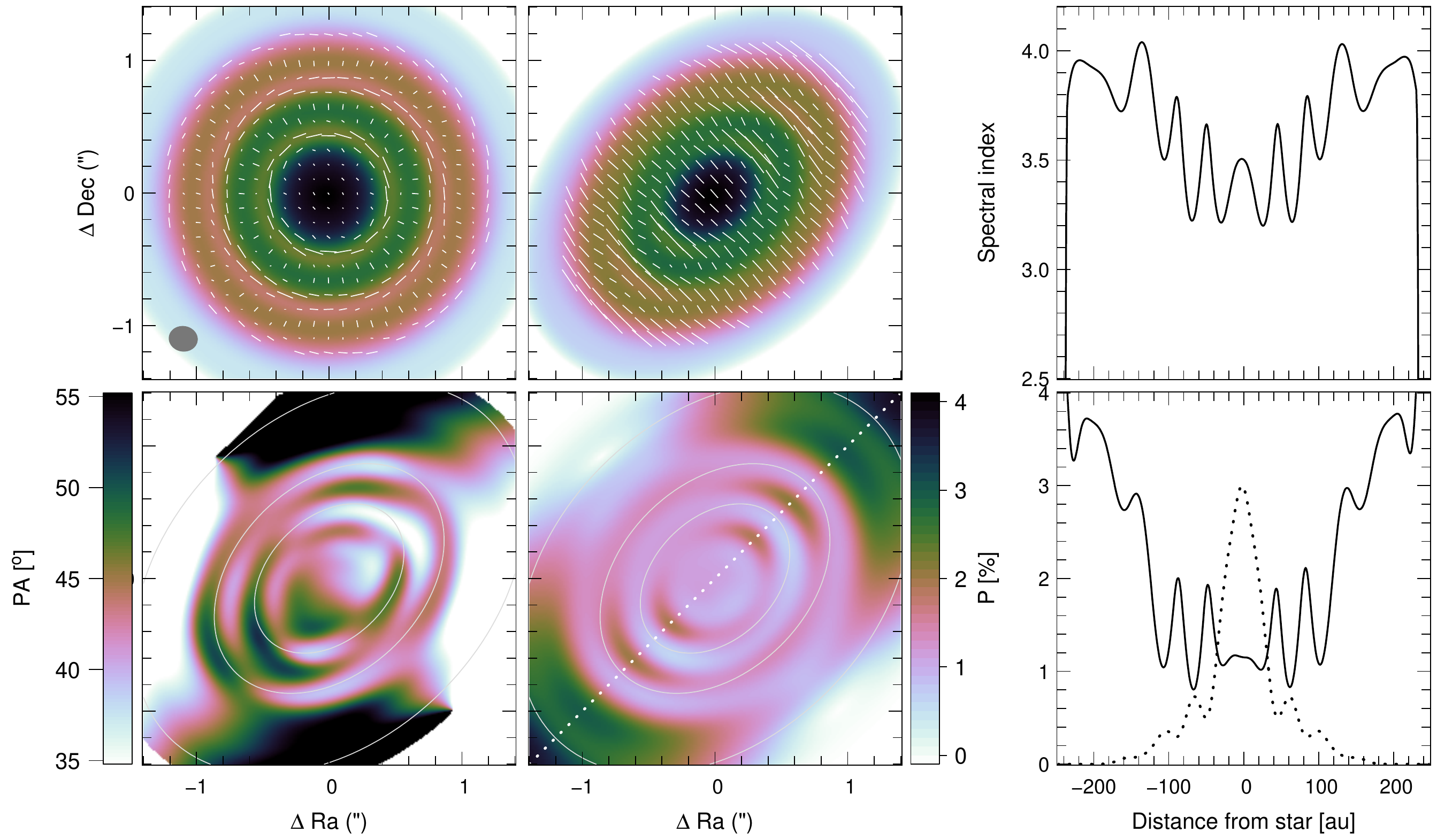}
  \caption{Sub-millimetre polarisation produced by self-scattering in our
      {\sf mcfost} model. \emph{Top left:} pole-on model, the background image is
    the intensity convolved by the observed beam with polarisation
    vectors overlaid. \emph{Top centre:} same model at a 45$\degr$
    inclination. \emph{Lower left \& centre:} polarisation angle and fraction. White ellipses mark
    the location of the bright intensity rings. 1$\arcsec$ corresponds to 105$au$. \emph{Top right:} spectral index; \emph{lower right:} polarisation along the major axis, with intensity (arbitrary units) shown dotted. 
    \label{fig:model}}
\end{figure*}

To simulate scattering in a ringed system, we use the radiative transfer code {\sc mcfost}  \citep{Pinte2006,Pinte2009}, building a disk density model by inverting the intensity image with uniform \textsc{clean} weighting into a surface density
profile. We assume a passive disk with a 28$L_{\odot}$ central stellar heating source and a Gaussian scale height of
  10\,au at 100\,au \citep{deGregorio2013}. Details are outlined in \cite{Pinte2016} for
HL~Tau. Spherically symmetric dust grains are used, where the only source of polarisation is
scattering (multiple scattering is included). The corresponding Stokes Q and U maps are computed,
and then convolved to the observed resolution (Fig.~\ref{fig:model}).

Polarization by scattering can reproduce the general features observed in the ALMA
data. In particular:
\begin{itemize}
  \item The polarisation is larger along the major axis,
    \emph{i.e.,} where the scattering angle is closer to 90$^{\circ}$.
\item The fractional polarisation is anti-correlated with the intensity, with larger
    polarization inside the gaps (darker areas in Fig.4, lower centre panel) \citep{Kataoka2016a}.
  \item The polarization vectors are aligned more with the disk minor axis in the central region and the intensity rings.
 \item The polarization angle displays a twist in the
   gaps, becoming more azimuthal (Fig.4, left centre panel). 
\end{itemize}

The twist in polarisation angle in the gaps reflects the change in the
scattering geometry and is more easily understood by looking at the same
model viewed pole-on (Fig.~\ref{fig:model}, top left). In the rings, the
radiation field is mostly azimuthal, producing a radial polarisation, while
in the gaps, the radiation field is mostly radial, originating from the
surrounding rings, resulting in an azimuthal polarisation pattern \citep[e.g.,][]{Kataoka2015,Kataoka2016a,Yang2016a}.
The disks around CW Tau \citep{Bacciotti2018} and GGD27 \citep{Girart2018} do not (yet) have resolved rings, but both show polarisation alignment along the minor axis in the centre, with a more azimuthal orientation at larger radii. This was interpreted as a change in scattering geometry as the disks become optically thin further out. HD163296 has the same basic geometry, except there are multiple transitions between the optically thicker rings and optically thin gaps.

The polarised intensity is strongly dependent on the chosen grain model, and is affected by a combination of the dust albedo, scattering phase
function and polarizability phase function. We calculate dust optical
properties using a distribution of hollow spheres (DHS) with a
maximum void fraction of 0.8. The model presented in Fig.~\ref{fig:model}
corresponds to compact astrosilicates \citep{Draine2003}, with a maximum grain size
a$_{max} \sim$\,100$\,\mu$m, typical of that required to maximise the
polarisation \cite[e.g.,][]{Kataoka2015,Kataoka2016a}.
This gives $\alpha_{mm}$\,$\sim$\,3.5 (larger in the optically-thinner gaps, top right of Fig.4), marginally consistent with HD163296 beyond a radius of $125\,au$. The low $\alpha_{mm}$ at smaller radii may be explained by large (mm-sized) grains, but this is inconsistent with a$_{max}$ from the polarisation \citep{Yang2016b}.  This dichotomy may be explained if the inner rings are optically thick but radially unresolved. Alternatively a$_{max}$ may depend on the vertical (or radial) location, with the 100$\,\mu$m grains on the surface (or in the gaps) and the mm dust in the midplane (or in the rings). In either case, the scattering geometry would not significantly change.
Further dust modeling, together with observations with higher resolution and longer wavelengths may be able to reconcile the low $\alpha_{mm}$ with the observed high polarisation fraction.

\section*{Acknowledgements}

This paper makes use of the following ALMA data: ADS/JAO.ALMA\#2015.0.00616.S and \#2013.0.00601.S. ALMA is a partnership of ESO (representing its member states), NSF (USA) and NINS (Japan), together with NRC (Canada), MOST and ASIAA (Taiwan), and KASI (Republic of Korea), in cooperation with the Republic of Chile. The Joint ALMA Observatory is operated by ESO,
AUI/NRAO and NAOJ. NRAO is a
facility of the NSF operated under cooperative
agreement by AUI. CP acknowledges funding from the Australian Research Council via
FT170100040 and DP18010423. FM acknowledges funding from ANR of France via contract ANR-16-CE31-0013. Thanks to Satoshi Ohashi for spotting an error in an earlier version of this paper.




\bibliographystyle{mnras}
\bibliography{HD-refs} 

\bsp	
\label{lastpage}
\end{document}


\section{Appendix: polarisation images}

The images of the polarisation Stokes components from the band 7 data are given in the figure. These images are obtained using {\sc clean} with natural uv weighting. The Stokes Q and U images show notably different structures compared with the total intensity (note that the intensity stretch in Figure A1 is linear for Q, U and V compared with logarithmic for I).

A non-zero V component can be seen in the upper right panel in V; this is detected at a level of $\sim$5$\sigma$, or $\sim 0.2\%$ of the peak flux. It is correlated more with the total intensity rather than with the polarised ring emission, and at present it is unclear whether this is real or a low-level calibration artifact. Currently, the formal ALMA accuracy of 0.5\% prevent us from making strong conclusions about the HD163296 circular polarization.

\begin{figure*}
   \centering
    \subfloat{
        \includegraphics[width=0.43\textwidth]{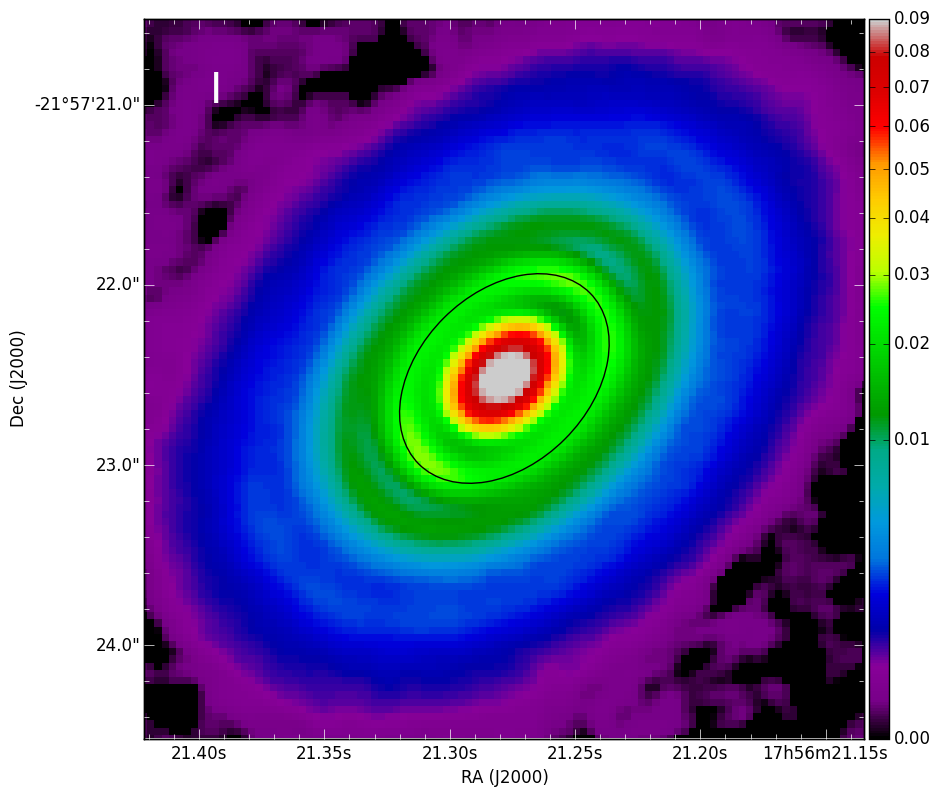}}
   	\qquad
    \subfloat{
        \includegraphics[width=0.45\textwidth]{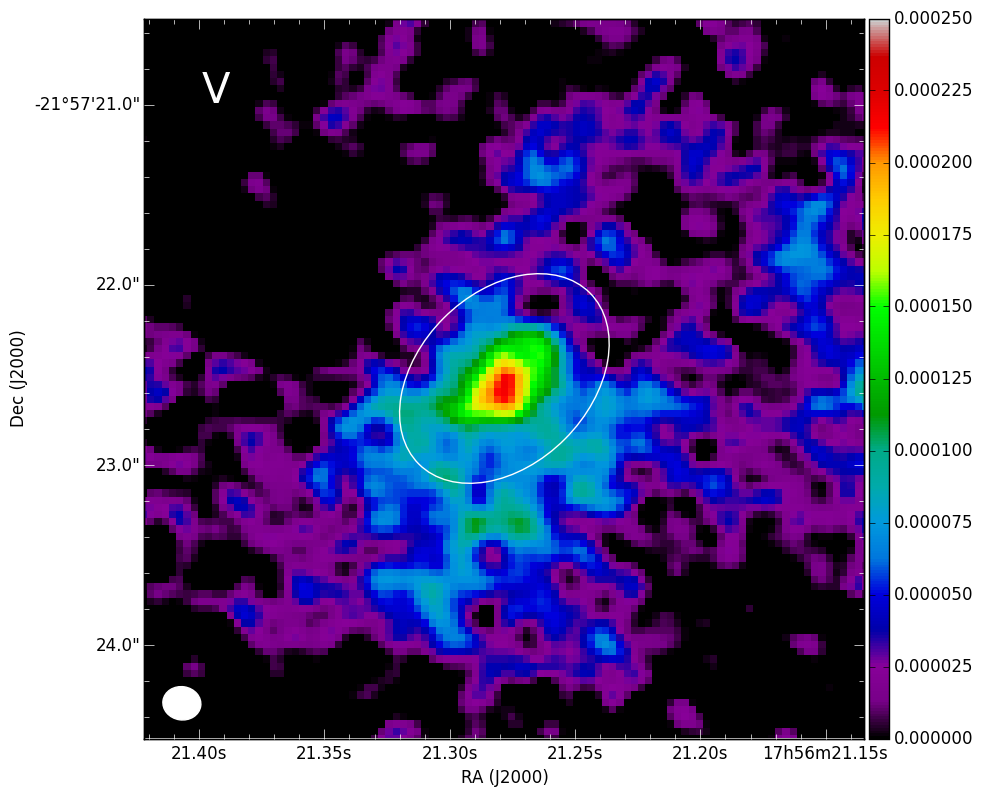}}
    
        \subfloat{
        \includegraphics[width=0.46\textwidth]{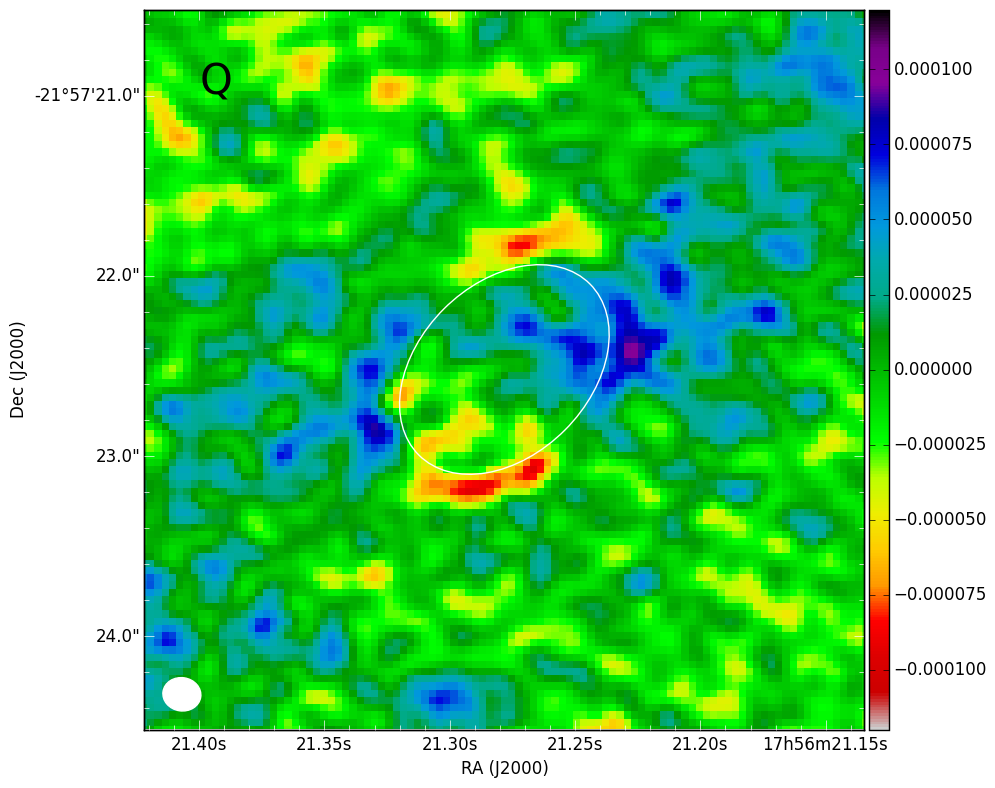}}
        \qquad
    \subfloat{
        \includegraphics[width=0.45\textwidth]{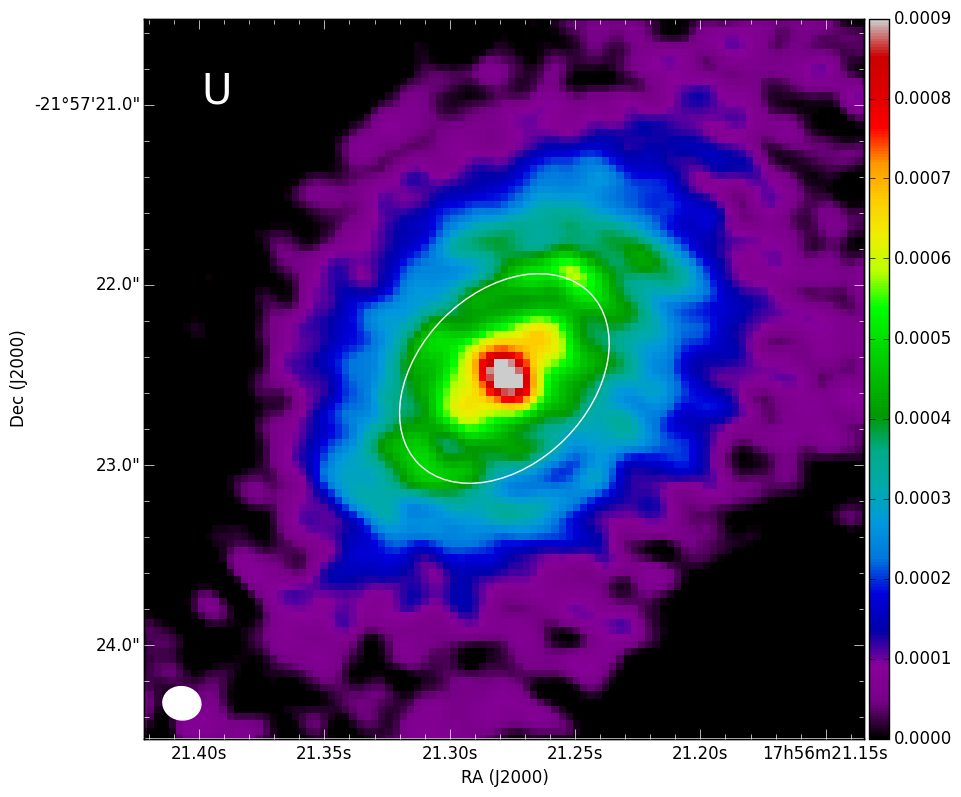}}

\caption{Polarisation images of HD163296 at band 7. Top panels are Stokes I and V components, and lower panels are Stokes Q and U. Intensity scales have a linear stretch for all except the I image. The rms in each of the datasets is $\sim$50$\mu Jy$. Each image is 4.0 arcsec across and the beam size is shown in the lower left. An ellipse representing the inner ring (1) is shown on each panel.}
    \label{fig:HD_4_allStokes.png}
\end{figure*}
